\documentclass[11pt]{amsart}
\usepackage{amsfonts,amssymb,amscd,amsmath,enumerate,verbatim,calc}
\usepackage[colorlinks,linkcolor=blue,pagebackref=true, citecolor=red]{hyperref}
\usepackage[all]{xy}
\usepackage{multicol}
\usepackage[mathscr]{euscript} 
\usepackage{mathrsfs}
\usepackage[left=2cm,right=2cm,top=2.5cm,bottom=2.5cm]{geometry}

\usepackage{graphicx}
\usepackage{color}
\usepackage{tikz}
\usepackage{tikz-cd}

\theoremstyle{plain}

\theoremstyle{definition}

\newtheorem*{example*}{\it Example}

\theoremstyle{remark}
\newtheorem*{claim*}{\it Claim}

\newtheorem*{case*}{\it Case}
\newtheorem*{note*}{\it Note}

\title[]{NEURALOGRAM: A  NEURAL NETWORK BASED REPRESENTATION \\ [2mm] FOR UNDERSTANDING AUDIO SIGNALS}
\thanks{\it The authors are affiliated with the Center for Computer Research in Music and Acoustics, Stanford University, CA}

\author{Prateek Verma}
\author{Chris Chafe}
\author{Jonathan Berger}  
\address{CCRMA, Stanford University Stanford, CA}
\email{prateekv@ccrma.stanford.edu   }
\email{cc@ccrma.stanford.edu  }
\email{brg@ccrma.stanford.edu   }

\begin{document}
\thispagestyle{empty}

\begin{abstract}
We propose the \textit{Neuralogram} 
\footnote{The term \textit{Neuralogram} elides \textit{Neural} based representation and\textit{ spectrogram}}- 
a deep neural network based representation for understanding audio signals which, as the name suggests, transforms an audio signal to a dense, compact representation based upon embeddings learned via a neural architecture. Through a series of probing signals, we show how our representation can encapsulate pitch, timbre and rhythm-based information, and other attributes. This representation suggests a method for revealing meaningful relationships in arbitrarily long audio signals that are not readily represented by existing algorithms. This has the potential for numerous applications in audio understanding, music recommendation, meta-data extraction to name a few.
\end{abstract}

\maketitle

\section{Introduction and Related Work}
\label{sec:intro}
Deep neural networks \cite{deeplearningbook} have had an enormous impact in various fields including audio signal processing. These algorithms have led to exploration of new fundamental approaches to classical problems in audio processing such as speech recognition \cite{las}, text-to-speech synthesis \cite{wavenet}, audio transforms \cite{haque2018conditional}, music generation \cite{dlfm}, sound understanding \cite{freqvermaschafer,CNNarch, anand2015convoluted}, and unsupervised audio processing \cite{verma2018neural}. Most of the problems in which they have been successful can be reduced to mapping a set of fixed or variable length vectors to a single or variable length output. These abstractions can be generalized across domains for a variety of applications. We propose ways in which we may use the power of deep neural architectures for devising new audio representations.  It presents opportunities for novel transforms, analogous to how the Fourier transform  enabled the short-time Fourier Transform (STFT) and  numerous variants (for example, the constant-Q transform \cite{brown1991calculation}, chromagram \cite{bartsch2001catch}, correlogram \cite{duda1990correlograms, licklider1951duplex}, RASTA \cite{hermansky1994rasta}) that subsequently emerged. These variants can be approximated as linear transformations applied to the power spectrum of the STFT. The newly devised representations were often formulated to overcome shortcomings existing in previously known representations and for use in a particular application. While linear spectrograms can capture the frequency content across time, they do not represent human pitch perception, and are purely a mathematical formulation. The constant-Q transform mimics the fact that the spacing between frequencies is constant in the log scale, in order to mimic human hearing. Similar arguments can be provided for the chromagram, which folds all frequencies into a fixed-dimensional vector. Rhyth-mogram or correlogram is another similar transform, created by stacking up successive computations of the autocorrelation function and then processing stacked representation
for various applications \cite{verma2015structural, paulus2009music,jensen2007multiple, engel2017neural}. The impact of deep learning in fields like natural language processing has been due to learning better representations for words by a fixed dimensional  vector. This vector can understand semantic dependencies, structure of different words, and learn the dependencies that exist with its neighborhood words. Recently the importance of learning a better embedding vector for words was shown \cite{peters2018deep}. It played a significant role in improving the performance of a variety of language processing tasks over existing methods. There have been several variants of algorithms for natural language processing that have been adapted in speech \cite{chung2018speech2vec,sentenceembedding}, graph networks \cite{grover2016node2vec} to name a few, which shows generalization of these approaches. \par
    In this work we introduce a new  representation of arbitrarily long audio signals based on a deep neural architecture. The main contributions of the paper are:

    1. A novel representation based on embeddings extracted from large scale neural networks for audio signals. 

    2. We show how the technique can represent various attributes of the audio based on pitch, rhythm, and timbre.

\section{METHODOLOGY}
\label{sec:methhod}
\textit{Neuralograms} are stacked neural embeddings for representing audio signals, similar to stacked autocorrelation function for correlogram representation.
In order to compute a Neuralogram for a particular audio signal, we first start with learning an embedding associated with the input. Embedding vectors have been used widely in a plethora of applications ranging from audio, speech \cite{chung2018speech2vec}, language \cite{peters2018deep}, drawings \cite{ha2017neural}, paintings \cite{gatys2015neural} and music \cite{engel2017neural} . Depending on the manner in which they are computed, they can encapsulate various facets of the input, sometimes either compressing the input and/or reconstructing the original signal, extracting the features necessary for a particular application or giving a weak supervision to learn another domain \cite{aytar2016soundnet}. \newline
    For our case, our objective is twofold. First, this fixed dimensional vector should be able to learn a dense, compact representation of the input which can capture various attributes characteristic of an audio signal viz. pitch, timbre and rhythm. Secondly, it should also be able to extract a representation which is smaller than the original signal in temporal domain. We started with training a deep convolutional architecture on a large scale audio dataset. There has been work showing how the state of the art algorithms for image understanding can be applied to understanding audio, and the performance gains achieved in one domain can be applied to the other \cite{CNNarch}. We trained a 19-layer convolutional architecture similar to the VGG architecture \cite{simonyan2014very}. The goal is to first have a neural architecture capable of interpreting the acoustic content on a known audio representation, i.e. linear spectrogram, in order to capture small temporal frequency variations with high resolution. Contrary to almost all of the current work using mel-scale input \cite{CNNarch}, we used a linear spectrogram. The parameters of the spectrogram input are 10ms hop size, 30ms window with a signal of 2s, down sampled to 8kHz, yielding a representation of 129x200. There are modifications in the pooling strategies to account for different dimensions of the input signal in frequency and time. 

\begin{figure}
  \includegraphics[width=350pt,keepaspectratio]{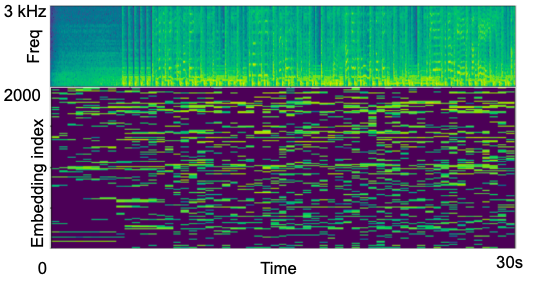}
  \caption{The spectrogram (top) and \textit{Neuralogram} of the  track "Happy" by "Pharell Williams" for the first 30s. Notice one can already draw associations of parts of spectrogram to neuralogram. The indices are different with Spectrogram being of the dimension 129x3000, having hop of 10ms and neuralogram of the size 500x58, having hop of 500ms  }
  \label{fig:fig1}
\end{figure}

Since the AudioSet labels have multiple output categories for a particular input, softmax was carried out for the last layer with the objective criteria to minimize the Euclidean distance between the output of the softmax layer and the desired labels. The categories were represented by encoding a 1 for the presence of the label and 0 otherwise in a fixed dimensional vector. The Adam optimizer \cite{kingma2014adam} was used as the optimization algorithm for adjusting the weights, with Xavier initialization \cite{glorot2010understanding}, and a dropout rate of 0.5 \cite{srivastava2014dropout} for regularization. The balanced subset of AudioSet was used in order to train the data.  However there do exist many ways to derive audio embeddings, and the current approach is just one of them. For example,  \cite{engel2017neural} explicitly provided pitch and used a waveform and a spectrogram-based autoencoder in order to get the embeddings for a musical note. On the other hand, \cite{haque2018conditional} used an architecture for transforming audio signal, and extracted embeddings for understanding the acoustic content.

Once the model is trained, we extract the last convolutional output in order to extract the embeddings of the input audio signal represented by a linear spectrogram. Similar to how we compute STFT, we slide a window across the input signal to get the embeddings at a particular time instant. Mathematically, STFT based spectrogram representation $X_{spec}$ and its variants can be described as below.

$${STFT} \{x[n]\}(m,\omega )\equiv X(m,\omega )=\sum _{n=-\infty }^{\infty }x[n]w[m-n]e^{-j\omega m}$$

$$X_{spec}\equiv |X(m ,\omega )|^{2}$$

Here $n$ denotes the time domain signal, with $m$ being the hop, and $\omega$ being the index corresponding to the frequency with the choice of window function $w$. Most of the variants  of spectrogram representation $\mathcal{L}{_{X_{spec}}}$ like constant-Q transform, chromagram, mel-spectrogram can be computed from a linear spectrogram via a simple linear transformation  i.e.,
\[
\mathcal{L}{_{X_{spec}}}= \mathcal{T}(X_{spec})
\]

Notice how a single layer neural net can essentially learn the transform $\mathcal{T}$ and perhaps much more depending on the problem of interest, thereby bypassing the gains that were achieved by hand-built custom transforms for specific applications over the past several decades. 

In order to draw a correspondence between \textit{Neuralogram} $X_{neur}$ and above representations we can describe it analytically as , 

$$X_{neur}(m,j )=\sum _{n=-\infty }^{\infty } \mathcal{E}^{l}_{k}((x[n]w[m-n])$$ where $\mathcal{E}^{l}_{k}$ represents the embeddings from the $l^{th}$ layer of a deep neural net $\mathcal{E}_{k}$ on raw waveform $x[n]$. As shown in \cite{sainath2015learning}, the magnitude spectrogram can be approximated as the output of the first layer of a convolution architecture. There are typically small relative performance gains achieved from going from raw waveform inputs to spectrogram inputs for convolutional architecture as shown in \cite{zhu2016learning}. \textit{Neuralogram} can also be interpreted as an audio signal being projected via a learned neural architecture to a fixed dimensional space (instead of projecting onto sinusoidal basis for traditional STFT) and these vectors can be  stacked across time. There can be many variants depending on the architecture of a neural net, and the layer from which the embeddings are extracted. The index of the embedding, denoted by $m$, ranges from 1 to size of the embedding $N$.

Notice that there do exist parallels to traditional signal processing based representations both in terms of parameter selection and the manner of computation. For neuralogram, we can choose how large the context input should be, (equivalent to window size in spectrogram), hop size or how densely we are sampling the input to the neural architecture, and the size of the embedding space (roughly equivalent to the resolution of the FFT). There can be various pros and cons for selecting parameters for construction of a neuralogram which may vary depending on the application of interest. 

A context window for the input to extract a single embedding, must be large enough to capture the signal of interest on which the deep architecture has been trained. On the other hand, it needs to be small enough to predict  within the window characteristics of the signal. Similar arguments can be given for size of the embedding space. We discuss the effects of embedding size in detail in section 3. It is important to note that all of the information will be shuffled across the  indices, so there is no correspondence between the index value and the desired attribute. The idea is to capture and extract the characteristics via manipulation of indices.
Since the indices of the \textit{Neuralogram} based representation do not necessarily correspond to a particular attribute, the representation is similar to word2vec \cite{mikolov2013distributed}. To give an example, training deep convolutional architecture with square filters on  \textit{Neuralogram} would yield to poor results, as there is no structure across the vertical dimension. A better model would be to use filters which are across \textit{all} the dimensions of the embedding space. In natural language processing, there have been applications in which the word2vec \cite{mikolov2013distributed} input representation was used, with model learning filters across the embedding space. The approaches learn fixed weights across the \textit{whole} embedding dimension in the first layer instead just parts of it. The deeper layers then combine the representation learned from the initial layers. There is a sufficient body of literature on handling shuffled input representation \cite{ha2017neural}, and tackling and developing models to handle such inputs, which may yield exciting approaches in future.

\section{EXPERIMENTS and discussion}
\label{sec:method}
The following sections describe a series of experiments to demonstrate the embedding-based representation and its interpretation, and how the \textit{Neuralogram} captures pitch, timbre and rhythmic characteristics - key attributes of a musical signal all with application to a wide range of audio understanding. We show that the representation learns features orthogonal to spectra (upon which it was trained), including  pitch and rhythm. 

\subsection{Understanding Pitch via probing chirp signals}
\label{ssec:pitch}
Trained to classify a large variety of spectral-timbral textures, the model was never explicitly given any information regarding musical pitch. We compute the \textit{Neuralogram} for a linear sinusoidal chirp starting from 4000Hz to 1Hz and then back again to 4000Hz. As mentioned in the previous section, there is no correspondence to the index of the embedding at a particular time instant to that of a particular attribute associated with the input signal. We learn a shuffling pattern with the assumption that we are only sorting via index of the pitch activations. A random signal representation would also yield the same activation pattern, but would lack the salience of the peak as seen in Fig. 2. Similar strategies have been deployed for filter-bank activation in \cite{freqvermaschafer,sainath2015learning}. Notably, there cease to be any activations in the lowest frequency regions. Only a subset of neurons gets fired. The remainder are dormant or have low relative activation for the entire frequency spectrum. We observe a linear activation associated with the chirp signal with a few filters getting activated for a particular frequency range in a linear fashion. This proves that the embeddings learn frequency content of the input signal implicitly. 

\begin{figure}
  \includegraphics[width=350pt,keepaspectratio]{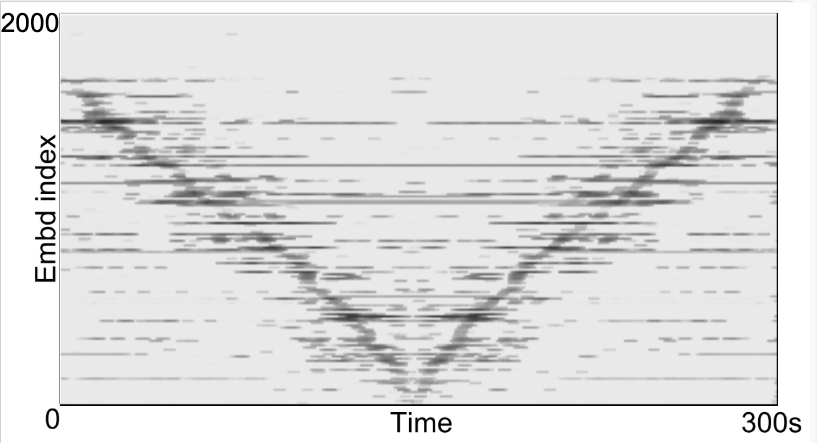}
  \caption{Neuralogram, with the index shuffled, (via index of frequency activations) for linear chirps first decreasing in frequency and increasing in frequency in the range from 4000Hz to 1Hz}
  \label{fig:sampel}
\end{figure}

\subsection{Understanding Timbre}
This section explains how our model is useful for understanding the timbre of a sound. Using the labels provided by the AudioSet corpus we are able to transform the input spectrogram to the desired timbral label via a deep neural architecture. There is a large body of research including \cite{CNNarch} on devising models to understand timbre. We must however note here that the current labels provided in AudioSet are inadequate and incomplete.  For example, a particular piece might have just a single label (eg \textit{pop song}). However, our trained architecture can produce categories much richer than the AudioSet label (eg, piano, drums, vocals, 'cello,\textit{pop song}) along with assigning a low probability to that of pop song. Conversely, where we predict a single correct label, there may be  multiple labels for a particular segment. Improving/reporting such numbers is not the goal of this work. (The performance of such models is investigated in \cite{CNNarch}). We do however note  that previous work, with similar architecture, have been shown to achieve state of the art performance for acoustic scene understanding, justifying the use of computing the current embedding and understanding the timbre of the signal. This was also reaffirmed by discovery of additional meta-data information for a particular audio clip which was not explicitly present in groundtruth Audioset labels.

\subsection{Effects of embedding size}

Figure 2 shows a \textit{Neuralogram} based representation for input of a linear embedding size of 2000. We tried changing the size of the embedding and found no significant degradation of the performance with respect to the  classification accuracy when we decreased the size from 2000 to 500. We presume that the depth of the network is sufficient to account for any changes in the embedding size, although a detailed study on very small embedding sizes needs to be carried out. The size of embeddings extracted was 128 in \cite{CNNarch}, and for future works this can be a hyper-parameter which will depend on the choice of application. The smaller the embedding size, the  more difficult it is to disentangle features and to a large extent, each of the attributes will correspond to multiple acoustic characteristics. 

\begin{figure}[b]
  \includegraphics[width=350pt,keepaspectratio]{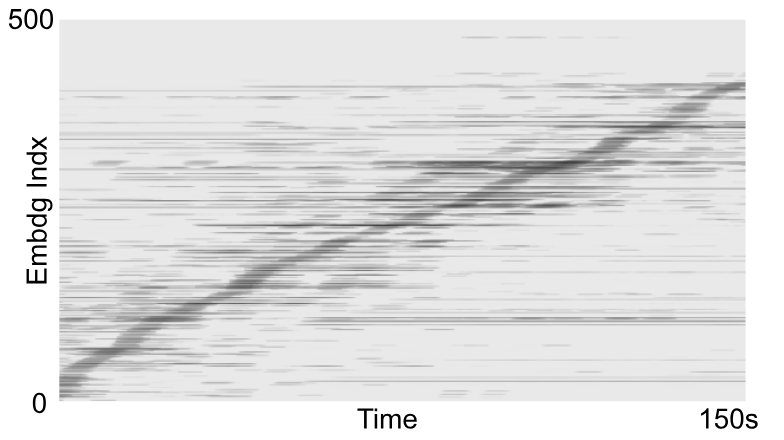}
  \caption{Neuralogram (sorted) for a smaller embedding size for linear chirp input from 1Hz to 4000Hz. Notice how coarse the activations are as compared to larger embedding size while retaining a linear response.}
  \label{fig:sampel}
\end{figure}

This also can be seen in Figure 3 which shows the response to a chirp signal, with similar shuffling strategies as Section 1. We see that the activation for a particular frequency no longer gives high salience in a few disjoint dimensions of the embedding. It results in bins giving strong responses resulting in a less narrow, multiple activation pattern.  The activation is also spread across multiple bins. But, it still retains similar characteristics of the higher size embedding.

\subsection{Understanding monotonic rhythmic pattern }
\label{ssec:pitch}

In order to observe whether the \textit{Neuralogram} can capture repetitiveness in the signal, we used an impulse train as an input. 
The impulse train had its period changing from 100ms to 1ms linearly over a period of 300s. This covers most of the rhythmic range that exists in audio signals that humans hear or  produce.  Figure 4 shows the result with the indices sorted, with highest activation of the higher periods being placed first. Note that the embedding emphasizes the repeated pattern present in the longer periods. As described in the previous section, the embeddings can still explicitly understand the higher frequency content. The cut-off happens around 20-30Hz irrespective of the starting period of the input signal. There can be several reasons behind this behaviour, one of them might be that the network is somehow learning to perceive sounds as we do although we have no evidence of that, yet. Almost all of the signals that have been used as an input, are sounds that humans listen to, and perhaps such a response, (and the response humans exhibits) is the optimal way of perceiving rhythmic content. In the region beyond the cut-off, we observed that it still exhibited sustained linear behaviour, but due to scaling and prominence of the activation before 20Hz, the activation strength is small. We also conducted a few experiments, keeping the rate fixed and changing the pattern or emphasis of beats and saw changes in the Neuralogram. 
There has been work on how important this boundary is in human hearing perception of rhythm and from perceived rhythm to pitched sound \cite{yost2009pitch,pressnitzer2001lower}. Our output also exhibits a piecewise linear behaviour and resembles closely to that of an optimal learned filterbank for a frequency estimation  architecture as shown in \cite{freqvermaschafer}.

\subsection{Understanding semantics}
We explored if, without any supervision, the semantic content of a signal can be understood from the network trained in previous section. This is different from all experiments above as the semantic content is not an acoustical property. The input given was multiple spoken words by multiple speakers from the speech commands dataset \cite{warden2018speech}. PCA and T-SNE maps were computed, and there was no observable separation of embeddings to various words or speakers. This might be due to fact that spoken language was categorized into speech, narration, conversation, female/male speech during training. Thus the features learned are only good for distinguishing these broad categories, and not finer nuances like the speaker and words. Fine-tuning or transfer-learning on these features can perhaps understand and separate these differences. Neural architectures having very similar characteristics with desired fine-grained labels that we have trained \cite{chung2018voxceleb2} have achieved state of the art results in understanding the contents of speech signals, and separation of speakers and/or words. 
   \begin{figure}
  \includegraphics[width=360pt]{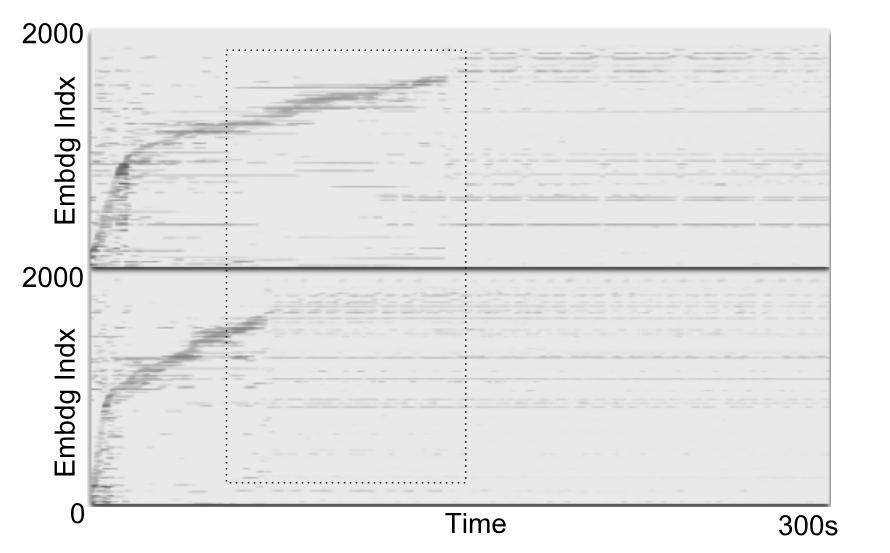}
  \caption{Neuralogram with indices sorted for an input of train of impulses starting from 100ms period (above) and 200ms period (below) ending at 1ms. Notice the piece-wise nature and abrupt ending at around 20-30Hz, (inside of the box) irrespective of the start point while retaining similar characteristics.}
  \label{fig:sampel}
\end{figure}

\section{Conclusion and Future Work}
We have demonstrated an audio transform based upon embeddings extracted from a large scale neural architecture that exposes meaningful qualities in audio. We have shown how this transform can encapsulate pitch, rhythm and timbral aspects of the audio signal. This has potential for numerous applications for understanding long audio signals. The correlations derived from the output shed new light into machine perception related to hearing, derived  from a data-driven analytical approach. We foresee many similar variants that may arise out of this representation.

\section{ACKNOWLEDGEMENTS}


\label{sec:acknowledgements}
The work was inspired by a lecture delivered by Malcolm Slaney, in CCRMA's \textit{Hearing Seminar}. The authors are grateful to him for his comments and suggestions. The authors also thank the Stanford Artificial Intelligence Laboratory for the use of their computing resources.

\bibliographystyle{IEEEbib}
\bibliography{references}

\end{document}